\documentclass[12pt]{article}
\usepackage[margin=1in]{geometry}
\usepackage[T1]{fontenc}
\usepackage[utf8]{inputenc}
\usepackage{amsmath,amssymb}
\usepackage{natbib}
\usepackage{hyperref}
\usepackage{url}
\usepackage{booktabs}
\usepackage{graphicx}
\usepackage{setspace}
\usepackage{placeins}
\hypersetup{colorlinks=true,linkcolor=blue,citecolor=blue,urlcolor=blue}
\usepackage[pagewise]{lineno}

\newcommand{\keywords}[1]{%
  \par\noindent\textbf{Keywords: }#1
}
\begin{document}

\begin{center}
{\Large\bfseries Coupled Supply and Demand Forecasting\\in Platform Accommodation Markets}\\[1em]
{\large Harrison E.\ Katz}\\[0.5em]
{Finance Data Science \& Strategy, Airbnb Inc., San Francisco, CA, USA}\\
{harrison.katz@airbnb.com}
\end{center}

\begin{abstract}
Tourism demand forecasting is methodologically mature, but it typically treats accommodation supply as fixed or exogenous. In platform-mediated short-term rentals, supply is elastic, decision-driven, and co-evolves with demand through pricing, information design, and interventions. Realized booked nights cannot exceed either latent demand or available supply, so booking models that ignore supply learn a regime-specific ceiling and become fragile under policy changes and supply shocks. This narrated review synthesizes work from tourism forecasting, revenue management, two-sided market economics, and Bayesian time-series methods; develops a three-part coupling framework spanning intervention, behavioral, and informational channels; and illustrates the identification failure with a toy simulation. I conclude with a focused research agenda for jointly forecasting supply, demand, and their compositions.
\end{abstract}

\keywords{Tourism forecasting; Two-sided markets; Compositional time series; Platform economics; Coupled systems; Censored demand; Supply forecasting}

% ============================================================
\section{Introduction}
\label{sec:intro}
% ============================================================

Tourism forecasting has a long tradition of modeling demand as a destination-level time series (arrivals, spending, hotel nights) with methods organized around econometric specifications, univariate time series, and a growing body of machine learning \citep{SongLi2008,SongQiuPark2019}. Two widely cited syntheses survey the post-2000 period and track the evolution of methods and evidence across five decades. This review is narrower and explicitly forward-looking: it targets forecasting in accommodation markets where supply and demand are coupled by design, especially platform-mediated short-term rentals.

The crucial shift is that ``capacity'' is no longer slow-moving and near-fixed (as in most hotel markets), but elastic at multiple margins: host entry and exit, listing activation and deactivation, and day-level availability decisions. Empirically, peer-to-peer accommodation supply is more flexible than hotel capacity, and the welfare gains from peer entry are largest in locations and periods where hotel capacity binds \citep{FarronatoFradkin2022}. Host entry, exit, and calendar activity also show that effective platform inventory varies at multiple margins \citep{FanEtAl2023,Wang2024scrape}. From a theory standpoint, two-sided market economics provides the backbone for why forecasting demand alone is incomplete in such settings: platforms internalize cross-side externalities via pricing, matching, governance, and information design \citep{RochetTirole2003,Weyl2010,Armstrong2006}.

A second motivation is that the post-2020 environment made structural breaks and regime change unavoidable in tourism and hospitality time series. That reality pushes modern forecasting toward methods that (i)~represent time variation and discontinuities, (ii)~remain probabilistically calibrated under intervention and drift, and (iii)~encode constraints (such as nonnegativity and shares summing to one) by construction rather than by post-hoc adjustment.

These concerns are not purely methodological. Destination managers, tourism boards, and policymakers rely on accommodation forecasts to allocate budgets, plan infrastructure, and evaluate regulatory interventions. When those forecasts are built on models that treat supply as fixed, they produce misleading signals: a city that introduces a short-term rental registration rule and then observes declining bookings may wrongly conclude that tourism demand has fallen, when in fact the binding constraint has shifted from demand to supply. The coupling framework developed here provides the conceptual tools to avoid such misdiagnosis.

The paper makes three contributions. First, it develops a coupling framework that distinguishes intervention coupling (rule or regime changes), behavioral coupling (host and guest responses to those changes), and informational coupling (the fact that bookings are observed under a moving and only partly observed supply ceiling). Second, it uses a stylized simulation to isolate one identification failure: after a supply shock, demand-only forecasters systematically over-predict because they have learned a regime-specific ceiling from the training period. Third, it outlines a five-problem research agenda whose resolution would constitute the core of a coupled forecasting research program for platform markets. A detailed survey of the methodological building blocks for coupled forecasting is provided in the Supplementary Material.

\subsection*{Notation and forecasting targets}

Platform accommodation data have a natural two-dimensional time structure: a \textit{stay date} $s$ (the night a guest occupies a listing) and a \textit{booking date} $b$ (the date the reservation is made), with lead time $\ell = s - b$. Throughout the paper, $t$ indexes stay-date periods (weeks or months) and all variables are aggregated over lead times unless otherwise noted.

Let $D_t$ denote latent demand intent: the number of listing-nights that guests would book for stay dates in period $t$ if inventory were unconstrained and matching were frictionless. Let $S_t$ denote effective supply: the number of listing-nights made available by hosts for stay dates in period $t$, defined as physical availability (active listing, unblocked night, not already reserved). Price ($P_t$) and platform visibility/ranking ($R_t$) are treated as separate covariates that influence both $D_t$ and $S_t$ but are not inventory in the same units as bookings. Let $B_t$ denote realized booked nights (the number of listing-nights that result in a completed reservation). Measurement issues affecting the construction of $S_t$, $B_t$, and demand proxies from external data sources are summarized below and expanded in the Supplementary Material.

Within a segment $k$ narrow enough that listings are close substitutes, realized booked nights satisfy
\begin{equation}\label{eq:segment}
B_{k,t} \leq \min(D_{k,t},\; S_{k,t}),
\end{equation}
with equality under perfect within-segment matching and no cross-segment substitution. Here $D_{k,t}$ denotes demand intent \textit{at prevailing prices and platform exposure}: it is not a deep structural primitive but the quantity of listing-nights that guests would book in segment $k$ given the prices $P_t$ and ranking $R_t$ they actually face, if inventory were unconstrained. Changes in $P_t$ or $R_t$ shift $D_{k,t}$; the inequality captures only the additional constraint imposed by physical availability $S_{k,t}$. Aggregating across segments, $B_t = \sum_{k} B_{k,t} \leq \sum_{k} \min(D_{k,t}, S_{k,t})$. The gap reflects matching frictions (search costs, preference heterogeneity, booking abandonment). A derived summary is the matching efficiency $m_t := B_t / \min(D_t, S_t) \leq 1$. Under perfect substitutability within the entire market, $m_t = 1$ and $B_t = \min(D_t, S_t)$. This stylized single-segment benchmark is used in several places for expositional clarity, with the understanding that matching frictions strengthen rather than weaken the coupling argument. Note that $m_t$ is a conceptual quantity: because $D_t$ is latent and $S_t$ is measured with error, $m_t$ is not directly observable and must be estimated from fitted model components or bounded from observable proxies.

This segment-level formulation also clarifies scope. The ceiling logic is strongest at the listing level, within narrow segments of close substitutes, and for short-run city forecasting after identifiable supply shocks. At broader aggregation or longer horizons, substitution across neighborhoods or property types, price adjustment, and host entry or re-entry can attenuate the constraint. The single-segment benchmark is therefore an analytic device, not a claim that every market-level series is equally ceiling-dominated.

The primary forecasting targets are:
\begin{enumerate}
\item \textit{Realized booked nights} $B_t$: directly observed; the primary operational forecast target. Measured in listing-nights (not reservation counts). Success metric: out-of-sample point accuracy (MAE, MASE) and distributional calibration (CRPS).
\item \textit{Latent demand} $D_t$: not directly observed; inferred from $B_t$ and supply proxies. The uncensored demand level is needed for counterfactual analysis, policy evaluation, and capacity planning. Success metric: structural adequacy and coherence with the observed $B_t$ constraint.
\item \textit{Available supply} $S_t$: partially observed and often noisy. Public calendar scrapes recover an imperfect proxy because blocked and booked nights are confounded; platform-internal aggregates are cleaner but usually unavailable to external researchers. When observables are listing-level (views, wishlists, dashboard states), market-level use requires explicit aggregation across listings or segments. Treated here as ``inventory'' when scalar and ``supply'' when referring to the broader distributional object (composition by host type, price tier, location). Success metric: out-of-sample accuracy and measurement fidelity.
\end{enumerate}

\subsection*{Measurement, observability, and aggregation}

Three empirical distinctions are important for the rest of the paper. First, public calendar scrapes and platform-internal records are different data environments. Public scrapes can approximate availability and occupancy changes, but they cannot cleanly distinguish blocked nights from booked nights. Internal records can do so directly. Second, several proposed observables, including listing views, wishlists, and dashboard metrics, are naturally listing-level objects. They become city-level or segment-level variables only if the platform aggregates them across listings. Third, scalar demand-intent proxies and distributional demand-intent proxies should not be conflated. Google Trends, card transactions, online reviews, and related public indicators can proxy aggregate demand intensity, but they do not recover booking lead-time distributions. The compositional version of the search-to-booking gap therefore requires internal search logs or equivalently granular search data.

\begin{table}[ht]
\centering
\caption{Data environments relevant for coupled forecasting in platform accommodation.}
\label{tab:data_env}
\small
\begin{tabular}{p{3.0cm}p{3.2cm}p{3.6cm}p{3.8cm}}
\toprule
\textbf{Data environment} & \textbf{Examples} & \textbf{What it can identify} & \textbf{Main limitation} \\
\midrule
Public scrape data & Repeated calendar snapshots; listing pages; vendor panels & Approximate availability, listing counts, occupancy changes, posted prices & Blocked and booked nights look identical; minimum-stay rules and coverage gaps create measurement error \\[0.6em]
Public aggregate demand proxies & Google Trends, card transactions, online review volume and valence & Scalar demand-intent signals and nowcasting covariates & Coarse aggregation; no lead-time composition; imperfect mapping to booking intent \\[0.6em]
Platform-internal aggregated data & Aggregated searches, views, wishlists, booked nights, exact calendar states & Cleaner measures of $B_t$ and $S_t$; lead-time-specific conversion gaps; market-level aggregation from listing data & Usually unavailable to external researchers \\[0.6em]
Administrative and regulatory data & Registration rolls, permit records, enforcement dates & Intervention timing, legal supply restrictions, external validation for supply shocks & Incomplete coverage of informal supply; often lagged relative to platform state \\
\bottomrule
\end{tabular}
\end{table}

The remainder of the paper is organized as follows. Section~\ref{sec:demand} reviews recent developments in tourism demand forecasting after the synthesis of \citet{SongQiuPark2019}. Section~\ref{sec:platform} reviews platform accommodation markets, two-sided foundations, and the supply-side forecasting gap. Section~\ref{sec:coupling} develops the coupling argument, clarifies the three coupling layers, and provides a stylized simulation-based demonstration. Section~\ref{sec:agenda} presents the research agenda with explicit identification assumptions, data environments, and evaluation designs. Section~\ref{sec:implications} discusses practitioner implications. Section~\ref{sec:limitations} states the boundary conditions and limitations of the framework. Section~\ref{sec:conclusion} concludes.

% ============================================================
\section{Recent developments in tourism demand forecasting}
\label{sec:demand}
% ============================================================

The recent tourism demand forecasting literature is better understood as the continuation of a mature forecasting toolkit under new data and new stress tests, rather than as a sequence of disconnected technical novelties. A useful recent stock-taking in \textit{Tourism Management} is \citet{SongQiuPark2023}, which revisits the field's theoretical and empirical development. For the present paper, three developments matter most: renewed attention to stabilization and combination, richer and faster data, and greater sensitivity to regime instability.

First, the literature continues to find that strong simple baselines remain competitive, which is why combination and stabilization have become central. \citet{Athanasopoulos2011} show why such baselines remain essential in tourism forecasting: automated time-series methods perform well, and na\"ive annual forecasts are hard to beat. Forecast combination has therefore remained both a practical baseline and an active research frontier \citep{BatesGranger1969,Timmermann2006}, with \citet{SongBagging2021} showing that bagging-based combination can stabilize accuracy in both normal and disrupted periods. Recent \textit{Tourism Management} work refines this point by showing that decomposition choices materially affect bagging performance \citep{LiuDecompBagging2023}. Related work on decomposition and nonlinear pattern recognition can be read in the same light: recurrent networks, Transformers, and compound pattern-recognition systems are attractive because they can absorb irregular seasonality and abrupt shifts \citep{ZhangDecomp2021,Salinas2020,Lim2021,HuCompound2025}. In a coupled accommodation market, however, even flexible forecasting architectures can be misled if the target series is censored by an omitted or poorly measured supply ceiling.

Second, the literature increasingly expands the information set beyond conventional macroeconomic covariates. Search intensity, internet-activity measures, card-transaction data, and online reviews now appear routinely as leading indicators in tourism forecasting \citep{SunEtAl2019,LiLawEtAl2021,HuOnlineReviews2022,GrauEscolano2026}. Recent \textit{Tourism Management} studies show that tourist-generated online reviews can improve destination forecasting \citep{HuOnlineReviews2022}. Mixed-frequency and multi-source designs that combine high-frequency digital traces with lower-frequency aggregate information can also materially improve forecasts during COVID-era recovery \citep{WuMixedData2023,HuMIDAS2025}. At the same time, recent critical assessment warns that search-query applications are highly sensitive to query design, preprocessing, and model specification \citep{MikulicBaumgartner2025}. From the perspective of this paper, the key point is not simply that these variables can improve accuracy. Rather, they widen the information set around evolving demand conditions. Some of them, especially search and web-traffic measures, may proxy demand interest earlier and at higher frequency than bookings do, while others help characterize shifts in market conditions that are only imperfectly reflected in booking data alone. This is especially useful when observed bookings may be censored by a moving supply ceiling.

Third, the pandemic made regime instability a first-order forecasting problem. The tourism forecasting competition tradition already emphasized disciplined benchmarking \citep{Athanasopoulos2011}, but COVID-19 accelerated scenario thinking, explicit judgmental adjustment, and the use of external information \citep{LiSongCOVID2022}. This matters for the coupling argument because the pandemic was not a pure demand shock. It simultaneously changed travel intent, host participation, regulatory conditions, and the composition of travel across origins, lead times, and listing types. The recent literature therefore illustrates a broader point: progress in tourism demand forecasting has been substantial, but the forecasting object is still usually framed as demand evolving under an implicit or only slowly varying supply environment.

That formulation remains appropriate when capacity is approximately fixed or the supply environment changes slowly relative to the forecast horizon. It becomes a substantive limitation in platform accommodation, where supply is elastic, decision-driven, and endogenous to the demand being forecast.

% ============================================================
\section{Platform accommodation markets and the supply-side gap}
\label{sec:platform}
% ============================================================

\subsection{Two-sided foundations}

A two-sided market is an economic environment in which an intermediary facilitates interactions between two distinct user groups and participation decisions are interdependent across groups \citep{RochetTirole2006}. The defining feature for forecasting is that platform decisions jointly shape both sides of the market. The demand-side participation level depends on expected supply-side participation and vice versa, so equilibrium quantities are determined by a fixed point rather than by independent supply and demand curves.

The modern definition emphasizes that two-sidedness is the empirical and theoretical relevance of the price structure: how charges are allocated across sides affects transaction volume, not only the overall price level \citep{RochetTirole2006}. In platform accommodation, coupling operates through both price instruments (commission splits, service fees, host-side pricing recommendations, subsidies) and non-price design choices (search ranking, cancellation policy, review systems, UX). The foundational models show that platforms face intertwined tasks of choosing an overall price level and how to split it across sides, with the profit-maximizing structure internalizing cross-group effects \citep{RochetTirole2003,Armstrong2006,Weyl2010}.

The operational extension makes the coupling dynamic. \citet{Cachon2017} show how state-dependent prices and wages improve capacity utilization in on-demand settings with self-scheduling providers. \citet{BaiSoTang2019} incorporate endogenous supply and demand with waiting-time effects. \citet{LianVanRyzin2021} model platform growth as evolving stocks of supply and demand, with optimal policies involving front-loaded subsidies that deliberately shift the coupling.

\subsection{Platform accommodation: pricing, supply dynamics, and regulation}

The platform accommodation literature most relevant to coupled forecasting centers on pricing dynamics, supply dynamics, and regulatory intervention. On competitive impacts, \citet{Zervas2017} use difference-in-differences to estimate effects of Airbnb expansion on hotel performance. On pricing, host heterogeneity and professionalization shape how Airbnb listings are priced across market states \citep{AbrateViglia2022,Casamatta2022}. This matters for the coupling argument because pricing is one channel through which demand shocks can feed back into host-side participation and availability. Recent \textit{Tourism Management} evidence from European Airbnb markets during the pandemic also shows that response-stringency measures and demand shifts propagated into listing prices, with substantial differences between commercial and private hosts \citep{MiloneGunterZekan2023}.

On supply dynamics, \citet{FanEtAl2023} model exit and transition dynamics of Airbnb listings, showing that supply composition evolves through multiple event types. \citet{FarronatoFradkin2022} estimate a structural model of competition between peer hosts and hotels and show that welfare gains from peer entry concentrate where hotel capacity is constrained. Regulation has created a large quasi-experimental literature, with studies of city-specific restrictions showing measurable supply impacts and heterogeneous host responses \citep{Koster2021,Duso2024,Robertson2024,Bibler2025}.

\subsection{The supply-side forecasting gap}

The limited depth of rigorous supply-side forecasting is itself a key finding. Accommodation supply has been modeled in simultaneous equation systems for hotel markets \citep{QuXuTan2002,TsaiEtAl2006}, but these treat supply as endogenous to explanation, not as a forecast target with out-of-sample evaluation. In the short-term rental sector, \citet{LeeLin2023} fit diffusion models to Airbnb listing data, one of the few studies that treats supply growth as an explicit forecasting object.

The measurement problem is equally important. Because platforms do not release granular activity data, studies rely on scraped calendars, listing snapshots, or proprietary vendors. \citet{Wang2024scrape} demonstrate how daily calendar scrapes can be translated into interpretable activity measures. This measurement layer is directly connected to forecasting feasibility, because supply forecasting requires consistent, high-frequency state variables that the field has not yet standardized.

% ============================================================
\section{The coupling problem}
\label{sec:coupling}
% ============================================================

The preceding sections document two facts. First, the demand forecasting literature is methodologically mature. Second, the supply side of accommodation markets remains thin as a forecasting target. This section argues that the separation of demand forecasting from supply forecasting is not merely an organizational gap. In platform-mediated accommodation markets, it produces forecasts that are fragile under policy changes, supply regime shifts, and platform intervention.

Using the notation from Section~\ref{sec:intro}, the core observation is that realized bookings are bounded: $B_{k,t} \leq \min(D_{k,t}, S_{k,t})$ (Equation~\ref{eq:segment}), with derived matching efficiency $m_t \leq 1$. A model that forecasts $B_t$ without conditioning on $S_t$ is fitting a constrained and friction-affected outcome without representing the constraint or the friction. Such a model may achieve adequate predictive performance during stable periods when the supply ceiling and matching efficiency are approximately constant. But it is not structurally adequate: it cannot answer counterfactual questions, it will fail when the supply regime shifts, and its parameters are not policy-invariant. The distinction between predictive adequacy (a model forecasts well in the current regime) and structural adequacy (a model remains valid under counterfactuals and regime change) is central to the argument that follows.

The coupling operates through three distinct mechanisms. Analytically, intervention coupling is the trigger, behavioral coupling is the propagation channel, and informational coupling is the observability problem. In practice these mechanisms interact recursively, but separating their roles clarifies why different forecasting failures arise.

\begin{figure}[ht]
\centering
\small
\setlength{\tabcolsep}{6pt}
\renewcommand{\arraystretch}{1.2}
\begin{tabular}{p{0.27\textwidth}c p{0.27\textwidth}c p{0.27\textwidth}}
\centering \fbox{\parbox{0.25\textwidth}{\centering \textbf{Intervention coupling}\\[0.3em]
Rule or regime changes\\
(regulation, ranking, commissions, policy redesign)}} &
$\rightarrow$ &
\centering \fbox{\parbox{0.25\textwidth}{\centering \textbf{Behavioral coupling}\\[0.3em]
Host and guest responses\\
(entry/exit, pricing, availability, booking and cancellation behavior)}} &
$\rightarrow$ &
\centering \fbox{\parbox{0.25\textwidth}{\centering \textbf{Informational coupling}\\[0.3em]
Observed bookings under a moving, partly observed ceiling\\
(search-to-booking divergence, stock-out censoring)}} \\
\end{tabular}

\vspace{0.6em}
\parbox{0.92\textwidth}{\centering \textit{Forecasting implication:} decoupled models transfer parameters across regimes, miss cross-side propagation, and misread booked nights as unconstrained demand.}
\caption{Conceptual ordering of the coupling framework. Intervention coupling changes the rules of the market, behavioral coupling transmits the change through host and guest responses, and informational coupling determines what the forecaster observes in bookings. The mechanisms interact recursively in practice, but the ordering clarifies their analytical roles.}
\label{fig:coupling_map}
\end{figure}

\subsection{Behavioral coupling: cross-side propagation}
\label{sec:behavioral}

Behavioral coupling refers to the host-side and guest-side responses that transmit a shock, rule change, or market signal across sides. The intervention itself is analytically distinct. A regulator or platform may change a rule, but behavioral coupling is the way hosts and guests respond through entry and exit, pricing, availability, search, booking, and cancellation behavior.

Two-sided market theory provides the mechanism. Because participation on one side depends on expected participation on the other, a shock that initially lands on one side rarely stays there \citep{RochetTirole2003}. Consider a well-documented example: Airbnb's introduction of a 48-hour full-refund grace period. \citet{JiaJinWagman2021} show that Airbnb's pro-guest cancellation-policy changes shifted both demand and supply on the platform, with heterogeneous effects across host types. That makes the case especially useful here: a guest-facing rule can propagate through prices, occupancy, and host participation rather than remaining confined to one side of the market.

The forecasting implication is that cross-side propagation is endogenous to the market state. A demand model trained before such responses occur will mis-forecast afterward, not only because the rule changed, but because the host and guest responses reshape the participation margin on both sides. Host exit and entry, calendar activation and blocking, price resetting, and cancellation responses are therefore better viewed as behavioral manifestations of coupling. By contrast, rule changes such as commission adjustments, ranking changes, or cancellation-policy redesign are intervention variables discussed in Section~\ref{sec:intervention}.

\subsection{Informational coupling: the missing ceiling}
\label{sec:informational}

The most fundamental form of coupling is mechanical: realized bookings cannot exceed available supply. Under the stylized benchmark, $B_t = \min(D_t, S_t)$. This ceiling is less problematic in traditional hotel markets, where room supply is physical, known, and approximately constant.

In platform accommodation markets, this ceiling is neither observable nor constant. The available inventory on any given night is the result of host-level decisions (listing activation, calendar availability, minimum-stay requirements) that aggregate into a time-varying, endogenous supply surface. Host pricing decisions do not change the inventory count $S_t$ as defined here (a listing-night is either available or not), but they do affect which portion of $S_t$ is demanded at prevailing prices; price is thus a covariate that modulates the demand-side conversion rate, not a component of the physical supply measure.

This creates a specific informational problem for demand forecasting. For accommodation-focused targets in markets and periods where inventory binds, the observed outcome is censored by available supply. The censoring structure is akin to a censored or unconstrained-demand problem: the analyst observes $B_t$ but not $D_t$ when $D_t > S_t$, and the censoring threshold $S_t$ is itself stochastic and endogenous \citep{Tobin1958,Talluri2004}. If the model does not include supply state variables, it is doing one of two things. Either it is implicitly assuming $S_t \to \infty$, in which case it will over-forecast in supply-tight markets. Or it is fitting to historically observed $B_t$ values that were already constrained by the prevailing supply environment, in which case it has learned the shape of the censoring mechanism without representing it, and it will fail when the supply regime changes.

The informational problem is compounded by a measurement problem. The demand side benefits from decades of convergence on well-defined targets: tourist arrivals, overnight stays, expenditures. No equivalent consensus exists for platform supply. Some studies count listings \citep{Zervas2017}, others measure spatial penetration \citep{Quattrone2018}, others estimate entry and exit flows \citep{FanEtAl2023,Kolleck2025}. The relevant supply forecast target is distributional: a characterization of available inventory by host type, location, price tier, and temporal availability pattern. Compositional data methods provide a natural starting point for this class of target, because they are built for constrained share vectors rather than unconstrained scalar series \citep{Aitchison1986,KatzBruschWeiss2024}. The forecasting literature has not yet framed the supply problem explicitly in those terms.

\subsection{Intervention coupling: endogenous regime change}
\label{sec:intervention}

Intervention coupling concerns rule or regime changes that alter the mapping between supply, demand, and observed bookings. The defining feature is not that behavior changes. Behavior also changes under many ordinary market shocks. The defining feature is that a platform or regulator changes the institutional environment itself, often in response to the current state of the market.

External regulation provides the most visible examples. City-level short-term rental regulations affect supply directly by constraining the feasible listing set \citep{Koster2021,Duso2024,Robertson2024,Bibler2025}. From a forecasting perspective, the critical observation is that the post-regulation relationship between supply and demand differs from the pre-regulation relationship. A demand model trained on pre-regulation data is not merely out-of-date; it is estimating parameters that describe a relationship that no longer holds.

Platform design changes constitute a second, less observable class of intervention. Search ranking, pricing recommendations, matching rules, fee structures, and cancellation-policy redesigns are periodically updated in response to observed market conditions. This creates a form of policy endogeneity familiar from the macroeconomics literature \citep{Lucas1976}: the historical relationship between supply and demand partly reflects past platform decisions, and those decisions are themselves functions of market state variables.

External shocks introduce a third variant when they force a regime change in the coupling relationship itself. The pandemic, in particular, redistributed demand across origins, listing types, and lead-time profiles rather than merely shifting aggregate levels \citep{KatzSavageColes2025,YangEtAl2021}. These are compositional changes, not merely level changes. Standard structural-break tools that detect level shifts in scalar time series miss this directionality and dimensionality. One recent proposal is \citet{Katz2026DirShift}, which develops a Bayesian Dirichlet ARMA model with a directional-shift intervention mechanism for compositional structural breaks.

For analytical clarity, it is useful to distinguish three types of structural breaks in coupled systems: (i)~\textit{demand shocks} that shift $D_t$ while leaving $S_t$ initially unchanged; (ii)~\textit{supply shocks} that shift $S_t$ while leaving $D_t$ initially unchanged; and (iii)~\textit{platform intervention shocks} that alter the coupling relationship between $D_t$ and $S_t$. Standard break-detection methods are best suited to type (i) and type (ii) breaks. Type (iii) breaks, which manifest as changes in coupling parameters rather than in either marginal series, require the multivariate extensions discussed in the research agenda.

\subsection{Synthesis}

Behavioral, informational, and intervention coupling are distinct mechanisms, but they compound. A platform or regulator introduces a rule change (intervention coupling). Hosts and guests adjust participation, pricing, availability, and booking behavior (behavioral coupling). The resulting booking series is then observed under a moving and only partly measured supply ceiling (informational coupling). The feedback is recursive, but the decomposition is useful because each layer implies a different empirical problem and a different forecasting failure mode.

I argue that in platform accommodation markets with elastic, decision-driven supply and frequent interventions, methods which treat supply as exogenous or absent face a structural limitation. The omitted variable ($S_t$) is correlated with demand drivers, and the resulting parameter estimates are biased for causal and counterfactual purposes. For short-horizon extrapolation in stable supply environments, the bias may be small and decoupled forecasts may perform adequately. But for any task that requires policy invariance, counterfactual reasoning, or robustness to supply regime shifts, the limitation is fundamental.

Table~\ref{tab:framework} summarizes the three coupling layers, their observable proxies, the failure modes they create for decoupled forecasters, and the corresponding research directions proposed in Section~\ref{sec:agenda}.

\begin{table}[ht]
\centering
\caption{Coupling framework: layers, proxies, failure modes, and research directions.}
\label{tab:framework}
\small
\begin{tabular}{p{2.4cm}p{3.0cm}p{3.7cm}p{3.4cm}}
\toprule
\textbf{Coupling layer} & \textbf{Observable proxies} & \textbf{Failure mode (decoupled model)} & \textbf{Research direction} \\
\midrule
Behavioral (Sec.~\ref{sec:behavioral}) & Host entry and exit; calendar activation or blocking; price resetting; booking and cancellation responses & Cross-side propagation is missed; model parameters fit one participation regime and are carried into another & Elasticity asymmetry diagnostic; state-dependent participation modeling \\\\[0.6em]
Informational (Sec.~\ref{sec:informational}) & Search-to-booking divergence; occupancy rates; calendar availability; blocked-booked sensitivity & Bookings are misread as unconstrained demand; stock-out censoring is omitted & Moving ceiling; search-to-booking gap; variance decomposition \\\\[0.6em]
Intervention (Sec.~\ref{sec:intervention}) & Regulation dates; ranking changes; commission changes; cancellation-policy reforms; A/B tests & Regime-specific parameters are applied across institutional regimes; Lucas-critique problem & Causal forecasting under unobservable intervention \\
\bottomrule
\end{tabular}
\end{table}

\subsection{A simulation-based demonstration}
\label{sec:simulation}

The following toy example is intentionally stylized. Its purpose is not to establish universal dominance of a coupled model over all possible demand-side benchmarks. It isolates one failure mode: if a model is trained in a regime where supply rarely binds, a later supply shock causes demand-only forecasts to extrapolate a ceiling that no longer exists. A market is observed over $T = 200$ periods with endogenous supply that responds to lagged demand. Realized bookings are $B_t = \min(D_t, S_t)$. In the first 150 periods, supply is sufficiently high that the constraint rarely binds. At $t = 151$, a regulation shock permanently reduces the supply intercept, simulating a registration requirement that removes casual hosts. Full model specification is given in the Supplementary Material.

A demand-only forecaster fits an AR(1) to $\{B_1, \ldots, B_{150}\}$ and extrapolates forward. Because the supply constraint rarely bound during training, the fitted model approximates the uncensored demand process. After the supply shock, the demand-only forecast continues to predict $B_t \approx 50$, while realized bookings drop to $B_t \approx 35$--$40$ because supply now binds. A coupled forecaster that conditions on $S_t$ detects the supply shift and adjusts.

\begin{figure}[ht]
\centering
\includegraphics[width=0.85\textwidth]{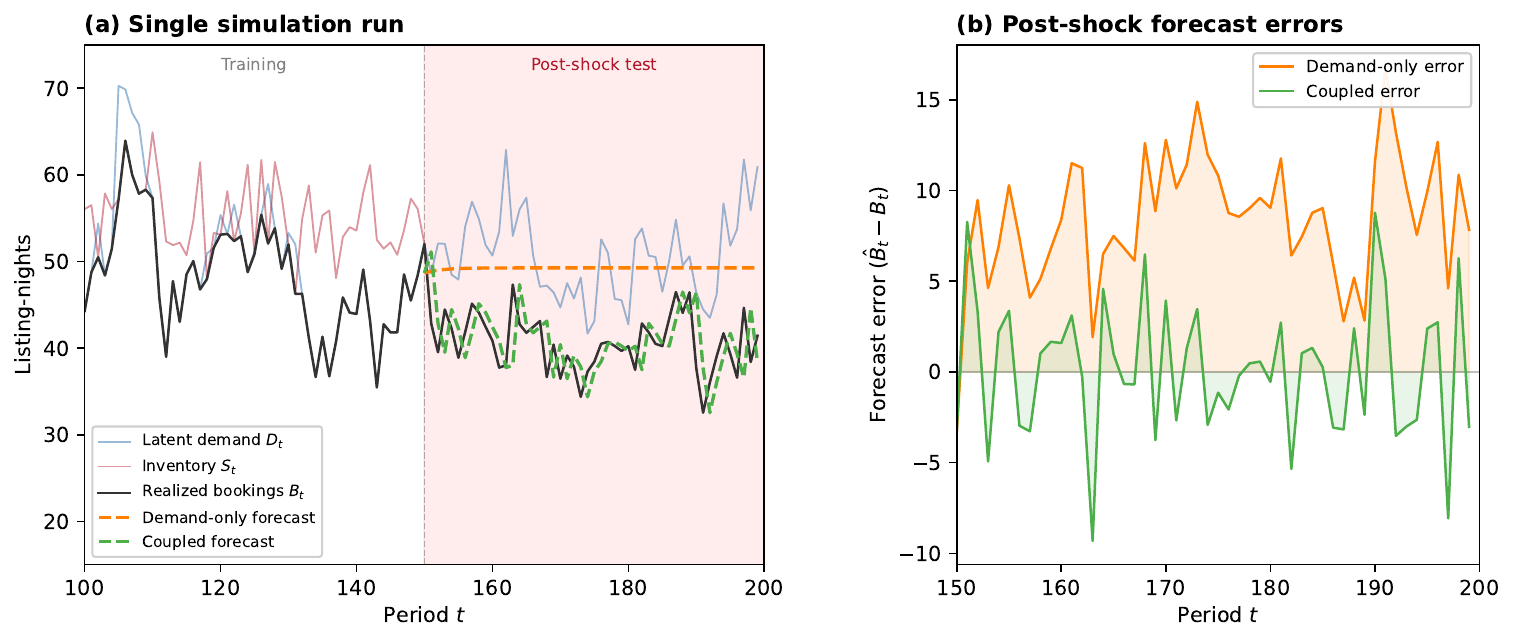}
\caption{Simulation-based demonstration. (a)~A single run: latent demand $D_t$, inventory $S_t$, realized bookings $B_t$, and the two forecast models over the test period. The supply intercept drops at $t = 151$. (b)~Post-shock forecast errors. The demand-only model exhibits a persistent positive bias; the coupled model tracks the new ceiling.}
\label{fig:sim}
\end{figure}

\begin{table}[ht]
\centering
\caption{Simulation results: demand-only vs.\ coupled forecaster (500 Monte Carlo replications).}
\label{tab:sim}
\begin{tabular}{lcccc}
\toprule
 & \multicolumn{2}{c}{Pre-shock ($t = 131$--$150$)} & \multicolumn{2}{c}{Post-shock ($t = 151$--$200$)} \\
\cmidrule(lr){2-3}\cmidrule(lr){4-5}
Model & RMSE & Mean bias & RMSE & Mean bias \\
\midrule
Demand-only AR(1) & 5.1 & $-0.3$ & 13.4 & $+11.8$ \\
Coupled $\min(\hat{D}_t, \hat{S}_t)$ & 5.2 & $-0.2$ & 5.4 & $+0.4$ \\
\bottomrule
\end{tabular}
\end{table}

Table~\ref{tab:sim} summarizes results across 500 Monte Carlo replications. The coupled model reduces post-shock RMSE by approximately 60\% and eliminates the systematic positive bias, while matching pre-shock accuracy. Two features merit emphasis. First, the demand-only model was \textit{accurate} during the training regime; the problem is not poor model fitting but fitted parameters that describe a relationship that no longer holds. Second, the failure is undetectable by standard residual diagnostics during training and manifests only out of sample when conditions change. This is the operational content of the distinction between predictive and structural adequacy.

Because the data-generating process is aligned with the coupled model and treats supply as directly observed, this exercise should be read as a mechanism demonstration rather than as a comprehensive horse race. The more demanding empirical question is whether supply-aware models outperform strong demand-only benchmarks when supply is noisy and only partially observed. Section~\ref{sec:min_design} therefore proposes an evaluation design with richer demand-only comparators and explicit sensitivity to supply-measurement error.

\FloatBarrier

% ============================================================
\section{Research agenda}
\label{sec:agenda}
% ============================================================

The coupling problem implies that forecasting in platform accommodation markets requires methods that jointly model supply and demand, respect endogenous constraints, and remain valid under intervention. While individual building blocks exist in adjacent literatures (revenue management, structural IO, marketplace design), they have not been assembled into an integrated forecasting framework for platform accommodation. The agenda below is organized in three layers. Sections~\ref{sec:search_gap} and 5.3 address foundational measurement and identification problems. Sections~5.4, 5.5, and~\ref{sec:min_design} describe near-term empirical designs that can be implemented with currently available public or proprietary data. Section~5.6 is the frontier problem: forecasting under unobserved intervention. Table~\ref{tab:agenda} summarizes the minimum data, identification route, and validation target for each problem. A more detailed technical survey is provided in the Supplementary Material.

\begin{table}[ht]
\centering
\caption{Research agenda for coupled forecasting in platform accommodation.}
\label{tab:agenda}
\small
\begin{tabular}{p{2.8cm}p{1.8cm}p{2.9cm}p{3.3cm}p{2.8cm}}
\toprule
\textbf{Research problem} & \textbf{Type} & \textbf{Minimum data} & \textbf{Plausible design} & \textbf{Validation target} \\
\midrule
Search-to-booking gap & Foundational & Bookings plus a public intent proxy for the scalar version; internal search logs for the compositional version & Scalar ratio or paired-composition distance used as a nowcasting or regime-state variable & Gap widens when stock-outs bind and improves forecast performance \\\\[0.5em]
Endogenous supply ceiling & Foundational & Bookings, supply proxy, and either an instrument or a documented supply shock & Constrained state-space or simultaneous-equation design with partial identification if needed & Lower post-shock bias and coherent latent-demand estimates \\\\[0.5em]
Variance decomposition & Applied & Bookings, supply states, and strong demand-side covariates & Nested forecast comparison between demand-only and coupled systems & Economically meaningful reduction in MAE, CRPS, or bias attributable to supply states \\\\[0.5em]
Elasticity asymmetry & Applied & Panel of prices, availability, bookings, and exogenous shifters & Time-varying supply and demand elasticities linked to forecast losses & High asymmetry predicts when decoupled models fail \\\\[0.5em]
Unobservable intervention & Frontier & Long multivariate panels with suspected latent breaks & Break-detection or robust-forecasting methods on the coupling relationship itself & Calibration is maintained across latent regime changes \\
\bottomrule
\end{tabular}
\end{table}

\subsection{Methodological landscape}

Five bodies of work bear directly on the coupling problem, each addressing a different facet.

\textit{Tourism demand forecasting} provides the baseline toolkit: econometric models, exponential smoothing, machine learning, and forecast combination \citep{SongQiuPark2019,SongLi2008,SongQiuPark2023}. This literature excels at modeling demand as a standalone time series and has converged on rigorous evaluation protocols. Its limitation, from the coupling perspective, is that supply enters either as a fixed background condition or not at all. When the supply ceiling is approximately constant, this omission is benign; when it shifts, the toolkit produces forecasts that are predictively adequate in-sample but structurally fragile out-of-sample (Section~\ref{sec:simulation}).

\textit{Revenue management and unconstrained demand estimation} addresses a closely related problem: recovering latent demand from sales data censored by finite capacity \citep{Talluri2004}. EM algorithms, booking-curve detruncation, and Bayesian data augmentation are well-developed for hotel settings where the capacity constraint is scalar and observable. The key assumption that breaks in platform markets is that the censoring threshold is fixed and known. When inventory is decision-driven and stochastic, these methods require extension to endogenous, high-dimensional supply surfaces (see Supplementary Material, Section~\ref{app:methods}).

\textit{Two-sided market economics} explains why the coupling exists. Foundational models \citep{RochetTirole2003,Armstrong2006,Weyl2010} show that equilibrium quantities in platform markets are determined by a fixed point across sides, not by independent supply and demand schedules. Operational extensions model dynamic capacity utilization \citep{Cachon2017}, endogenous entry with waiting-time effects \citep{BaiSoTang2019}, and growth-stage subsidies that deliberately shift the coupling \citep{LianVanRyzin2021}. These models are structural and explanatory; they have not been operationalized as forecasting systems, but they provide the economic logic that a coupled forecasting framework must encode.

\textit{Causal and intervention inference} supplies the tools for evaluating what happens when coupling relationships change. Bayesian structural time series \citep{Brodersen2015}, synthetic control \citep{Abadie2010,Abadie2015}, and difference-in-differences designs are standard for retrospective policy evaluation. The deeper connection is to the Lucas critique \citep{Lucas1976}: parameters estimated under one policy regime are not invariant to policy change, which is precisely what intervention coupling creates. Extending these tools from retrospective evaluation to prospective forecasting under unobserved intervention is the hardest open problem identified below.

\textit{Compositional time series methods} round out the toolkit. Many platform forecasting targets are compositional: supply shares by host type, demand shares by origin market, lead-time allocations across booking horizons \citep{Aitchison1986,KatzBruschWeiss2024}. The coupling argument established that shocks redistribute mass across components rather than simply shifting a scalar level. Methods that respect simplex constraints, model time-varying volatility, and detect directional structural breaks on the simplex are necessary building blocks for representing the redistributive channel through which coupling operates.

Each tradition solves part of the problem. Tourism forecasting models demand well but ignores the supply constraint. Revenue management handles the constraint but assumes it is fixed. Two-sided economics explains the coupling but does not forecast. Causal inference evaluates past interventions but does not project forward. Compositional methods represent distributional targets but have not been integrated with supply-demand coupling. The remainder of this section identifies five open problems whose resolution would connect these building blocks into an integrated framework.

\subsection{The search-to-booking gap as a supply constraint signal}
\label{sec:search_gap}

In accommodation platforms, guests search before they book, and the temporal profile of search activity differs from that of booking activity. The \textit{search-to-booking gap} is defined here as the divergence between demand-intent proxies and realized bookings. Two versions should be distinguished. The \textit{scalar} version compares an aggregate demand-intent proxy, such as Google Trends or card-based travel search activity, with booked nights. The \textit{compositional} version compares full lead-time, spatial, or segment distributions of search intent and realized bookings.

This distinction matters empirically. Search volume data are already used as leading indicators of aggregate arrivals \citep{SunEtAl2019,LiLawEtAl2021,HuMIDAS2025}, but the informational content of the conversion margin is usually discarded. Recent critical assessment of Google Trends and Baidu Index applications in \textit{Tourism Management} underscores that query selection, preprocessing, and model specification remain central concerns when search-based proxies are used for tourism demand forecasting \citep{MikulicBaumgartner2025}. From the perspective of this paper, a widening scalar gap can still be useful as a public-data indicator that demand intent is not converting into bookings.

The stronger compositional object requires more data. It can be quantified using paired lead-time distributions and compositional distance measures, but Google Trends does not recover such lead-time distributions by itself. The compositional version therefore requires platform-internal search logs or equivalently granular data in which searches can be linked to intended stay dates. The public-data and internal-data versions are both useful, but they answer different questions and should not be conflated.

\subsection{Demand forecasting under an endogenous supply ceiling}

Section~\ref{sec:informational} established that realized demand is bounded by available supply and that this ceiling is time-varying and endogenous. The open problem is to build a demand forecasting framework that explicitly incorporates this constraint while being honest about measurement.

This has precedent. In labor economics, \citet{BarnichonNekarda2012} forecast the unemployment rate by modeling the flows that govern the stock. The simultaneous-equation tradition \citep{Haavelmo1943,ZellnerTheil1962} provides identification strategies for jointly determined systems. A promising direction is a two-equation or state-space system in which the demand equation conditions on a latent supply state and the supply equation conditions on a demand signal, with the constraint that realized bookings are bounded by both. Under platform-internal data, the supply state may be modeled directly from exact calendar and booking states. Under public data, the same problem is better viewed as a proxy or partial-identification problem because the supply variable is measured with error.

The central identification challenge is that $D_t$ and $S_t$ are not separately identified from $B_t \leq \min(D_t, S_t)$ without instruments or structural assumptions. Candidate instrument classes include: (i)~regulation shocks that constrain supply; (ii)~host-side fee or commission changes; (iii)~weather or events in origin markets that shift demand; and (iv)~timing assumptions exploiting the fact that supply decisions precede demand realization for a given stay date. Each class requires exclusion restrictions that are plausibly violated on platforms. The honest assessment is that clean instruments are rare in this setting, so research designs should state exclusion restrictions explicitly, propose falsification tests, and consider partial-identification bounds \citep{AtheyImbens2017}.

A further complication is supply measurement error. With public calendar scrapes, blocked and booked nights are confounded, minimum-stay rules distort night-level availability, and coverage varies across vendors. Practical designs should therefore either (i)~model $S_t$ as a latent state observed through noisy proxies, or (ii)~report sensitivity across alternative scrape-based construction rules and interval bounds. In this setting, an empirical success is not perfect identification. It is a coupled system that reduces post-shock bias, remains coherent with the booking constraint, and is robust to plausible supply-measurement perturbations.

\subsection{Decomposing demand variance into exogenous and endogenous components}

In a coupled system with fast feedback, observed demand variance has two components: genuine exogenous uncertainty and endogenous oscillation from coupling. A useful empirical design is therefore nested forecast comparison. Let $M^D$ denote a strong demand-only benchmark that uses the best available public demand-side covariates. Let $M^{DS}$ augment that system with supply states or a censoring layer. The variance attributable to coupling is then the reduction in forecast loss or predictive variance when moving from $M^D$ to $M^{DS}$, evaluated over rolling windows and especially around supply shocks.

This idea connects naturally to time-varying volatility models \citep{KatzWeiss2025BDARCH}, but the validation criterion should be forecasting, not just model fit. The relevant question is whether adding supply states produces economically meaningful improvements in MAE, MASE, CRPS, post-shock bias, or calibration. With public data, the exercise should be repeated across alternative supply-construction rules so that the variance reduction is reported as robust, attenuated, or fragile to measurement assumptions. That design would quantify whether coupling is merely conceptually plausible or actually a material source of recoverable forecast error.

\subsection{Price elasticity asymmetry as a coupling diagnostic}

Not all markets are equally coupled. When supply and demand respond to price signals with comparable elasticity, the system equilibrates more quickly and independent forecasts may suffice. When one side is substantially more elastic, persistent imbalances arise and coupling effects should be strongest. This suggests a diagnostic rather than a universal claim.

A workable research design would estimate time-varying supply and demand elasticities for a panel of platform markets using prices, availability, bookings, and exogenous shifters on each side. The object of interest is an asymmetry index, not the elasticity estimates in isolation. The empirical test is then whether periods and markets with high asymmetry also exhibit poor relative performance of decoupled forecasts, measured for example by \citet{DieboldMariano1995} comparisons against coupled alternatives. Validation comes from forecast loss, not just from elasticity significance. This would turn the coupling claim into a conditional statement: coupling matters most where the market fails to re-equilibrate quickly.

\subsection{Causal forecasting under unobservable platform intervention}

The most challenging problem is also the most consequential. Section~\ref{sec:intervention} argued that platforms intervene endogenously. Standard causal tools (difference-in-differences, synthetic control \citep{Abadie2010,Abadie2015}, Bayesian structural time series \citep{Brodersen2015}) require knowing when the intervention occurred. Structural break detection \citep{BaiPerron2003,Killick2012} can identify when the process changed but cannot distinguish exogenous shocks from endogenous platform responses.

Two directions are promising. First, break detection methods operating on the coupling relationship itself, parameterized through time-varying elasticities or the correlation structure of residuals, rather than individual series. Bayesian change-point methods \citep{Fearnhead2006,AdamsMacKay2007} provide the inferential machinery but require extension to multivariate, coupled settings with compositional constraints. Second, forecasting methods explicitly robust to unobserved intervention: bounding forecast uncertainty to account for possible coupling changes, related to conformal prediction \citep{Romano2019,Barber2021} but under the non-stationarity that endogenous intervention creates.

\subsection{Summary and prioritization}

The five problems are not independent, but they do have a logical progression. The search-to-booking gap and the moving ceiling are foundational measurement and identification problems. Variance decomposition and elasticity asymmetry are near-term empirical designs that can be attempted with current data if the measurement problem is handled carefully. Causal forecasting under unobservable intervention is the frontier problem because it requires methods that remain calibrated when the coupling relationship itself changes.

A researcher entering this agenda would do well to begin with the supply-measurement problem that underlies several formulations. Without canonical supply state variables that can be measured consistently across markets, the ceiling cannot be estimated, the variance cannot be decomposed, and the coupling cannot be parameterized. The payoff is substantial. Methods developed for platform accommodation will generalize to other platform markets (ride-sharing, labor platforms, e-commerce marketplaces) where similar coupling dynamics are present but less well documented.

\subsection{A feasible empirical design and benchmark set}
\label{sec:min_design}

To make the agenda concrete, this subsection sketches a feasible empirical test of the coupling hypothesis. The design requires one city with a well-documented supply shock, one public demand-intent proxy, repeated supply measures, and a benchmark set that is stronger than a single univariate AR model.

\textit{Setting.} A city that introduced a short-term rental registration requirement with a known effective date. Candidate cities include San Francisco, Berlin, and Bordeaux, each of which has a documented regulation episode in the literature \citep{Bibler2025,Duso2024,Robertson2024}. Amsterdam remains useful as a modeling case, but \citet{OverwaterYorkeSmith2022} is an agent-based simulation rather than a documented quasi-experimental regulation study.

\textit{Supply construction.} $S_t$ is constructed from daily calendar scrapes as the count of available listing-nights for a 30-day forward stay-date window, following the transparent scrape-based pipeline of \citet{Wang2024scrape}. Because public scrapes cannot cleanly distinguish blocked from booked nights, the analysis should report sensitivity across alternative supply-construction rules or interval bounds rather than rely on a single point estimate.

\textit{Demand-intent signals.} Publicly available scalar intent proxies can be drawn from Google Trends \citep{GoogleTrends}, card-transaction data \citep{GrauEscolano2026}, and online review data \citep{HuOnlineReviews2022}. These variables can support demand-only nowcasting and provide the public-data version of the search-to-booking gap. The full lead-time compositional gap requires platform-internal search logs and is therefore a separate, stronger design.

\textit{Benchmark set.} The demand-only benchmark set should include at least: (i)~a univariate time-series model such as AR or ETS on $B_t$; (ii)~a demand-only model with public leading indicators; and (iii)~a mixed-frequency demand-only model when high-frequency search or review data are available. The coupled model should then be compared against this set, not only against the weakest benchmark.

\textit{Evaluation.} The coupled system forecasts $\hat{B}_t = \min(\hat{D}_t, \hat{S}_t)$, where $\hat{D}_t$ is allowed the same demand-side information set as the strongest demand-only benchmark and $\hat{S}_t$ is an inventory nowcast. Accuracy should be compared out of sample over the post-regulation period using MAE, MASE, post-shock bias, and where possible CRPS or interval calibration. If the coupled model significantly outperforms strong demand-only models after the supply shock but not before, this constitutes evidence that coupling matters when regimes change.

\textit{Data access.} The public-scrape version establishes a replicable baseline. If proprietary platform data are available, the same design can be executed with exact booking counts, exact calendar states, lead-time-tagged searches, and richer inventory signals. The public and proprietary implementations should be presented as nested versions of the same empirical design, not as interchangeable data environments.

% ============================================================
\section{Implications for destination management and tourism policy}
\label{sec:implications}
% ============================================================

The coupling framework has direct implications for practitioners who rely on accommodation forecasts for planning and resource allocation.

\textit{Destination managers and tourism boards.} The practical task is to distinguish demand weakness from a supply contraction. After a regulation change or platform shock, managers should track booked nights, available inventory, and at least one external demand-intent proxy together. A decline in bookings accompanied by stable demand intent and falling inventory should be treated as a supply shock, not as evidence that destination demand has collapsed. Marketing budgets, staffing plans, and infrastructure decisions should therefore be conditioned on the joint movement of bookings and inventory rather than on bookings alone.

\textit{Regulatory impact assessment.} Cities considering short-term rental regulation should not infer welfare or tourism effects from raw booking changes alone. A useful post-policy dashboard would separately monitor permit or registration counts, active inventory, host composition, prices, and booked nights. That decomposition allows analysts to distinguish supply withdrawal, price adjustment, and booking conversion effects. Regulatory impact assessments based only on pre-regulation demand trends risk overstating lost tourism revenue because they confound demand changes with mechanically reduced supply.

\textit{Platform operators.} Forecasting systems that support earnings guidance, treasury operations, and capacity planning should treat major ranking, commission, pricing-policy, or cancellation-policy changes as regime events. In practice this means tagging model vintages by intervention regime, monitoring forecast accuracy conditional on recent interventions, and retraining or switching models when the platform changes the rules of the market. Supply state variables should enter the forecasting system directly, and forecast monitoring should be stratified by inventory tightness rather than assessed only in aggregate.

\textit{Researchers using public data.} Public-data implementations remain useful, but they should be framed honestly as proxy-based versions of the coupled problem. Calendar scrapes provide noisy supply proxies, and Google Trends provides only scalar intent signals. A defensible public-data application therefore emphasizes sensitivity analysis, benchmark strength, and clear separation between what is identified directly and what is only inferred indirectly.

% ============================================================
\section{Boundary conditions and limitations}
\label{sec:limitations}
% ============================================================

The argument developed here is strongest at the level where the supply constraint is most local: individual listings, narrow segments of close substitutes, or short-run city forecasting around identifiable supply shocks. After broad aggregation or over longer horizons, substitution across neighborhoods and property types, price adjustment, and host entry or re-entry can partially restore market clearing. The claim is therefore not that every market-level booking series is always ceiling-dominated. It is that the risk becomes substantively important when supply changes faster than aggregation can smooth it.

A second limitation is empirical observability. Public calendar scrapes provide imperfect supply proxies because blocked and booked nights are confounded, minimum-stay rules distort night-level availability, and coverage varies across vendors. Listing views, wishlists, and many dashboard metrics are listing-level objects unless the platform aggregates them internally. Public demand-intent proxies such as Google Trends are useful scalar signals, but they do not identify lead-time compositions. Public-data implementations should therefore be treated as proxy-based or partially identified versions of the coupled problem.

A third limitation is that the simulation in Section~\ref{sec:simulation} is a stylized mechanism illustration. It isolates a supply-shock failure mode under observed supply and a deliberately simple benchmark. It does not by itself establish the incremental value of coupled forecasting relative to the strongest possible demand-only models with rich leading indicators.

Finally, the research agenda remains prospective. The paper identifies the data structures, identification challenges, and validation standards needed for coupled forecasting, but it does not yet deliver a fully estimated production-ready system. The contribution is therefore conceptual and programmatic, with Section~\ref{sec:min_design} intended as a concrete next step rather than as a completed empirical test.

% ============================================================
\section{Conclusion}
\label{sec:conclusion}
% ============================================================

The central argument of this paper is that in platform-mediated accommodation markets with elastic supply and frequent intervention, supply and demand cannot be forecasted independently when the goal is policy-invariant, counterfactual-capable, or structurally robust prediction. The capacity ceiling is time-varying and endogenous to demand. Platform interventions alter the structural relationship between the two sides. Behavioral responses to policy changes propagate across sides through cross-group externalities. Decoupled forecasts may perform adequately for short-horizon extrapolation in stable environments, but they are fragile precisely when forecasts matter most: under policy changes, supply regime shifts, and platform intervention.

The tourism demand forecasting literature has produced a mature and sophisticated toolkit for predicting arrivals, spending, and stays. This toolkit was developed for a world in which supply was effectively fixed and the censoring mechanism was approximately constant. In platform accommodation, the censoring threshold itself is elastic, decision-driven, and coupled to demand by design. I argue that the methods work; the problem formulation needs updating.

The research agenda outlined in Section~\ref{sec:agenda} offers a path forward. Its core is a conceptual reframing: from forecasting demand given fixed capacity to jointly modeling a coupled system in which both sides co-evolve under endogenous intervention. The methodological building blocks exist but have not been assembled for this purpose. The five open problems identified here constitute the work plan for that assembly.

While this paper focuses on platform accommodation, the coupling problem generalizes. Ride-sharing platforms face analogous supply-demand feedback through surge pricing and driver repositioning. Labor platforms mediate elastic, decision-driven supply with endogenous matching interventions. E-commerce marketplaces manage seller entry, inventory visibility, and algorithmic ranking that jointly shape both sides of the transaction. In each setting, forecasting one side without the other produces structurally fragile predictions. The accommodation market is a natural starting point because the data infrastructure, regulatory variation, and academic literature are most developed, but the methods proposed here are designed for the broader class of coupled platform markets.

% ============================================================
% SUPPLEMENTARY MATERIAL
% ============================================================
\clearpage
\appendix
\setcounter{section}{0}
\renewcommand{\thesection}{S\arabic{section}}
\renewcommand{\thetable}{S\arabic{table}}
\renewcommand{\thefigure}{S\arabic{figure}}
\renewcommand{\theequation}{S\arabic{equation}}

\begin{center}
{\Large\bfseries Supplementary Material}
\end{center}

\vspace{1em}
\noindent This supplement provides technical details supporting the main text. Appendix~\ref{app:measurement} expands on the measurement discussion already summarized in the main text. Appendix~\ref{app:simulation} gives the full simulation specification. Appendix~\ref{app:methods} surveys the methodological building blocks for coupled forecasting. Appendix~\ref{app:distance} formalizes the search-to-booking distance metrics.

% ============================================================
\section{Measurement and data sources}
\label{app:measurement}
% ============================================================

This appendix expands on the main-text discussion of measurement, observability, and aggregation. Inventory $S_t$ is typically constructed from calendar scrapes or platform dashboards. The primary measurement problems are: (i)~distinguishing blocked dates (host-side unavailability) from booked dates (guest-side demand), since both appear as ``not available'' in a calendar scrape; (ii)~handling minimum-stay rules, which make individual nights conditionally available depending on booking length; (iii)~accounting for multi-night bookings, which occupy consecutive listing-nights but represent a single demand event; and (iv)~sampling bias, since scrape frequency and coverage vary across vendors, and inactive or stealth listings inflate raw counts \citep{Wang2024scrape}. The conceptual object $S_t$ is defined as inventory available at any posted price; $P_t$ and $R_t$ then filter which portion of that inventory is exposed to demand under specific search conditions.

Demand intent proxies (searches, listing views, wishlists) are available on platform dashboards but generally not in public data. Credit-card aggregates and Google Trends provide partial substitutes at coarser granularity \citep{GrauEscolano2026,SunEtAl2019}. The gap between these proxies and the conceptual $D_t$ is itself informative (see the research agenda in the main text).

For external researchers, an honest assessment of data availability is: $S_t$ can be constructed from calendar scrapes, but with measurement error from the blocked-versus-booked ambiguity; $B_t$ can be approximated from occupancy changes, but not perfectly; $D_t$ is latent even with platform-internal data, because search intent is not equivalent to booking intent. $P_t$ and $R_t$ are partially observable from listing pages but may not reflect the prices or rankings actually displayed to searchers. This measurement gap is itself a research problem (see the main text).

% ============================================================
\section{Simulation specification}
\label{app:simulation}
% ============================================================

The simulation in Section~4 of the main text uses the following data-generating process. Latent demand follows an AR(1) process: $D_t = 50 + 0.7(D_{t-1} - 50) + \varepsilon^D_t$, with $\varepsilon^D_t \sim N(0, 5^2)$. Supply is endogenous and responds to lagged demand with noise: $S_t = 40 + 0.3 D_{t-1} + \varepsilon^S_t$, with $\varepsilon^S_t \sim N(0, 3^2)$. Realized bookings are $B_t = \min(D_t, S_t)$. In the first 150 periods (training), supply is sufficiently high that the constraint rarely binds. At $t = 151$, a regulation shock permanently reduces the supply intercept from 40 to 25, simulating a registration requirement that removes a portion of casual hosts.

The two forecasters are defined as follows.

\textit{Demand-only:} fit an AR(1) to $\{B_1, \ldots, B_{150}\}$ and produce one-step-ahead forecasts $\hat{B}_{t} = \hat{\mu} + \hat{\phi}(B_{t-1} - \hat{\mu})$.

\textit{Coupled:} fit an AR(1) to $\{B_1, \ldots, B_{150}\}$ for a latent demand proxy $\hat{D}_t$, and separately fit a regression $S_t = \alpha + \beta D_{t-1} + \eta_t$ to the observable supply series; then forecast $\hat{B}_t = \min(\hat{D}_t, \hat{S}_t)$. In the simulation, both $S_t$ and $B_t$ are observed; in practice, $S_t$ would be a noisy proxy from calendar scrapes.

Results are reported in Table~3 of the main text (500 Monte Carlo replications). The demand-only model achieves pre-shock RMSE of 5.1 and post-shock RMSE of 13.4 with mean bias $+11.8$. The coupled model achieves pre-shock RMSE of 5.2 and post-shock RMSE of 5.4 with mean bias $+0.4$, a roughly 60\% reduction in post-shock RMSE with elimination of the systematic positive bias.

% ============================================================
\section{Methodological building blocks}
\label{app:methods}
% ============================================================

This appendix surveys the methods most relevant to constructing a coupled forecasting framework. The building blocks span five areas: compositional and constrained forecasting, unconstrained demand estimation and censoring, causal inference under intervention, regime-change detection, and uncertainty quantification. None of these was developed specifically for the platform accommodation problem, but together they constitute the toolkit from which a coupled forecasting framework can be assembled.

\subsection{Compositional and constrained forecasting}

Many quantities relevant to coupled forecasting are compositional: market shares by guest origin, supply shares by host type, lead-time allocations across booking horizons, and currency-mix shares all must sum to one and lie on the simplex. The simplex constraint implies that raw shares are negatively correlated by construction (an increase in one component mechanically decreases others), which motivates log-ratio transformations that map compositions to unconstrained Euclidean space. Standard time series methods that ignore these constraints can produce incoherent forecasts: predicted shares that exceed unity or take negative values \citep{Aitchison1986,PawlowskyGlahn2015}.

Two core modeling strategies have emerged. The first is transform-based: apply a log-ratio transformation (additive, centered, or isometric) to map the simplex to unconstrained Euclidean space \citep{Egozcue2003}, then fit standard multivariate models. Early work by \citet{BrunsdonSmith1998} demonstrated the viability of ARIMA modeling on log-ratio-transformed compositions, and \citet{Kynclova2015} extended this to VAR models for multivariate compositional series. \citet{BarceloVidal2011} develop compositional VARIMA models within the Aitchison geometry, providing a general autoregressive-moving-average framework on the simplex. \citet{SnyderEtAl2017} develop a state-space formulation for compositional time series using the log-ratio approach, and \citet{Mills2010} provides a practical overview of forecasting methods for compositional data. Transform-based methods benefit from the full arsenal of Euclidean time series tools, but they can obscure interpretability and may perform poorly when shares approach zero \citep{CoendersFerrer2020}. The zero problem is non-trivial in practice; \citet{MartinFernandez2012} develop model-based replacement strategies for rounded zeros, and \citet{MartinFernandez2015} propose Bayesian-multiplicative treatments for count zeros in compositional data.

The second strategy models compositions directly on the simplex. The Bayesian tradition is particularly well developed. \citet{QuintanaWest1988} introduce dynamic linear models for compositional time series using a logistic-normal formulation, and \citet{Cargnoni1997} extend this to multinomial forecasting through conditionally Gaussian dynamic models, demonstrating posterior inference for market-share evolution. \citet{Grunwald1993} propose Bayesian state-space models for continuous proportions using the Dirichlet distribution, and \citet{ZhengChen2017} develop frequentist Dirichlet ARMA models. The logistic-normal distribution \citep{Aitchison1986,AitchisonShen1980} accommodates richer covariance structures on the simplex, at the cost of a less tractable normalizing constant. Dynamic multinomial logit models treat compositional observations as arising from category-level utility processes with time-varying parameters, connecting naturally to discrete-choice demand models in the \citet{BLP1995} tradition. \citet{Ravishanker2001} apply compositional time series methods to mortality proportions, illustrating the broader applicability of these techniques to demographic and social science data. For count-based compositions, Dirichlet-multinomial models provide overdispersed multinomial inference with conjugate updating \citep{PawlowskyGlahn2015}. A practical concern in applied settings is whether to model shares alone or jointly with their totals; \citet{CoendersEtAl2017} address this by incorporating a compositional predictor with a total in generalized linear models. Each distributional family involves different tradeoffs: Dirichlet models enforce simplex constraints automatically but impose limited covariance flexibility; logistic-normal models offer richer dependence but lose conjugacy; transform-based approaches are computationally convenient but can behave poorly at the simplex boundary.

Within the Dirichlet family, Bayesian Dirichlet ARMA (BDARMA) models combine the Dirichlet likelihood with VARMA dynamics on the mean parameters after an additive log-ratio transformation \citep{KatzBruschWeiss2024}. Extensions address time-varying volatility through a Dirichlet ARCH component \citep{KatzWeiss2025BDARCH}, structural breaks via a directional-shift intervention mechanism \citep{Katz2026DirShift}, and density forecast calibration through a centered-innovation formulation \citep{Katz2025centered}. Software implementation is available in the \texttt{darma} R package \citep{darmaR2024}.

For coupled forecasting, compositional modeling is often necessary for coherence and interpretability because the supply forecast target is itself distributional. The composition of available inventory by host type, price tier, or location is a simplex-valued object that evolves over time. The coupling argument established that supply constraints and platform interventions redistribute demand across components (origins, lead times, listing types) rather than simply shifting a scalar level. A forecasting framework that cannot represent redistributive shocks on the simplex will miss the primary mechanism through which coupling operates.

\subsection{Unconstrained demand estimation and demand censoring}

The censoring formulation $B_t \leq \min(D_t, S_t)$ connects the platform forecasting problem to a mature literature on unconstrained demand estimation in revenue management and demand censoring in inventory systems. In hotel revenue management, observed room sales are censored by available capacity, and recovering the latent demand distribution is a prerequisite for optimal pricing and allocation \citep{Talluri2004}. The statistical problem is classical: the forecaster observes a truncated or censored outcome and must infer the parameters of the uncensored distribution. Standard approaches include expectation-maximization algorithms for censored Poisson or normal demand \citep{Talluri2004}, booking curve detruncation methods that project partial booking curves to their uncensored completions \citep{doi:10.1177/004728759703600102,Weatherford1992}, and Bayesian formulations that place priors on both the demand parameters and the censoring probability.

The platform accommodation setting introduces three complications that go beyond the classical problem. First, the censoring threshold $S_t$ is not fixed or known; it is time-varying, endogenous to demand signals, and must itself be estimated or forecast. Second, the censoring is high-dimensional: supply constraints bind differently across listing types, locations, price tiers, and dates, so the relevant censoring mechanism operates on a compositional supply surface rather than a scalar capacity limit. Third, the censoring threshold responds to the same market conditions that drive demand, creating simultaneity that standard censored-demand estimators do not address. A concrete platform-specific example illustrates the difficulty: a host who observes strong demand may block weekend dates for personal use, or impose a three-night minimum stay that eliminates short-trip demand. Hotels also use close-outs and length-of-stay controls, but these decisions are centralized within the firm and typically observable to its revenue management system. In peer-to-peer platforms, thousands of heterogeneous hosts adjust availability rules idiosyncratically, making the effective censoring threshold stochastic and substantially harder to measure or forecast than in the hotel setting. These complications constitute the primary novelty of the platform censoring problem relative to the hotel revenue management baseline. A further practical difficulty is that external scrape-based supply proxies are noisy: calendar scrapes typically cannot distinguish ``booked'' from ``blocked'' nights without inference (see Appendix~\ref{app:measurement}), so any estimate of $S_t$ from public data carries measurement error that propagates into the censoring correction. Platform-internal data can separate booked and blocked states directly.

The connection is nonetheless valuable. Methodologically, it provides a well-understood starting point: the tools for censored likelihood estimation, EM algorithms for incomplete data, and Bayesian data augmentation for latent demand are directly applicable as building blocks. Conceptually, it clarifies what is and is not new in the platform setting. The existence of a supply constraint on observed demand is not new; the hotel literature has handled this for decades. What is new is the endogeneity, dimensionality, and decision-driven nature of the constraint itself.

\subsection{Causal inference meets forecasting}

Coupling is often policy-mediated, which makes causal inference tools directly relevant to forecasting in platform markets. The practical toolkit includes three families.

Bayesian structural time series (BSTS), operationalized in the CausalImpact framework \citep{Brodersen2015}, provide counterfactual prediction when intervention timing is known. BSTS models fit a state-space model to pre-intervention data and project it forward, using the discrepancy between the projection and the observed post-intervention outcome as an estimate of causal impact. This approach is natural for evaluating the effect of regulation or platform policy changes on accommodation metrics, but it requires a clean pre-period and known intervention timing, conditions that are often violated when platforms intervene continuously.

Synthetic control methods \citep{Abadie2010,Abadie2015} formalize counterfactual construction from a donor pool of untreated units. In the accommodation context, markets where regulation was introduced can be compared to synthetic control markets constructed from markets without the regulation. The method is powerful for single, discrete interventions but less suited to the continuous, overlapping interventions that characterize platform design changes.

Difference-in-differences designs are common in platform policy evaluation, including studies of cancellation policy effects \citep{JiaJinWagman2021} and regulation impacts \citep{Koster2021,Duso2024}. However, they inherit risks when treatment timing and heterogeneity interact with dynamic feedback, precisely the conditions that coupling creates. The broader literature on causality and policy evaluation \citep{AtheyImbens2017} provides the methodological standards that coupled forecasting must meet, including attention to heterogeneous treatment effects, staggered adoption, and interference between units.

For the research agenda, the gap is not the absence of causal tools but their integration with forecasting. Existing causal methods evaluate interventions retrospectively; the coupled forecasting problem requires causal reasoning prospectively, forecasting what will happen under interventions that have not yet occurred or that are unobservable.

\subsection{Regime change and structural breaks}

The post-2020 period made structural breaks unavoidable in tourism time series, and the accommodation sector was among the most severely affected. Point forecasts trained on stable regimes failed catastrophically, and the recovery exhibited heterogeneous trajectories across markets, segments, and metrics.

The classical approach to multiple structural breaks is the \citet{BaiPerron2003} framework, which estimates break timing and regime-specific parameters in regression structures. For algorithmic change-point detection with computational efficiency, the PELT algorithm \citep{Killick2012} provides optimal segmentation in linear time under certain penalty structures. Both approaches operate on scalar or low-dimensional time series and detect level or trend shifts.

Bayesian change-point models offer a probabilistic alternative. \citet{Fearnhead2006} develops exact Bayesian inference for multiple change-points, while \citet{AdamsMacKay2007} propose Bayesian online change-point detection (BOCPD) for sequential settings. The Bayesian framework is attractive for forecasting because it produces posterior probabilities over change-point locations, which can be propagated into prediction uncertainty. Tourism applications include structural break dating for arrivals series \citep{CroMartins2017} and Markov-switching models for demand cycles \citep{BothaSaayman2022}.

For platform accommodation, a key limitation of standard break detection is that it operates on scalar series and detects level shifts. The coupling argument implies that shocks redistribute mass across components rather than simply shifting a mean. The pandemic did not just reduce bookings; it shifted origin mixes, compressed lead-time distributions, altered host type compositions, and changed spatial patterns \citep{KatzSavageColes2025,YangEtAl2021}. These are compositional changes that require break detection on the simplex. The directional-shift DARMA framework \citep{Katz2026DirShift} addresses this gap by parameterizing a structural break as a direction of change on the simplex, an amplitude, and a logistic timing function, while maintaining the Dirichlet likelihood throughout.

\subsection{Adjacent approaches for joint modeling}

Several methodological traditions outside tourism forecasting address the problem of jointly modeling multiple endogenous variables and can serve as templates or components for coupled supply-demand forecasting.

\textit{Simultaneous equation models.} The econometric tradition of simultaneous equation estimation \citep{Haavelmo1943,ZellnerTheil1962} directly addresses the identification problem in jointly determined systems. Three-stage least squares extends single-equation instrumental variables to entire systems, exploiting cross-equation covariance and allowing cross-equation parameter restrictions.

\textit{Structural equilibrium models.} In industrial organization, \citet{BLP1995} develop an empirical framework for differentiated product markets that jointly models demand (discrete choice with consumer heterogeneity) and supply (firm first-order conditions from price competition). \citet{FarronatoFradkin2022} apply a related structural approach to the Airbnb--hotel competition setting.

\textit{Multi-task learning.} From the machine learning side, \citet{Caruana1997} frames multitask learning as inductive transfer: learning tasks in parallel with a shared representation can improve generalization because additional task signals act as a regularizing bias. Modern probabilistic deep forecasting models \citep{Salinas2020,Lim2021} implement multi-task logic at scale.

\textit{Agent-based models.} Agent-based modeling provides mechanism exploration through simulation of micro-level agent rules and the resulting macro-level dynamics \citep{LuxMarchesi1999,LeBaron2006}. In the platform accommodation context, \citet{Vinogradov2020} use ABM to simulate Airbnb supply growth under alternative regulatory regimes in Norwegian municipalities, and \citet{OverwaterYorkeSmith2022} model the interaction between short-term rentals, housing prices, and residential displacement in Amsterdam.

\textit{Joint forecasting in adjacent domains.} In electricity markets, \citet{Maciejowska2024} show that the largest accuracy gains appear when forecasting functions of jointly distributed quantities. In labor markets, \citet{BarnichonNekarda2012} forecast the unemployment rate by modeling the labor force flows that govern the stock, offering a transferable template for platform accommodation where bookings are the constrained intersection of supply availability and demand intent.

\subsection{Uncertainty quantification}

Uncertainty quantification is harder under coupling because forecast errors across targets are correlated and state-dependent.

For comparing point forecast accuracy, the \citet{DieboldMariano1995} framework remains standard. For interval forecasts, \citet{Christoffersen1998} conditional coverage tests detect clustered violations, an important diagnostic when coupling creates correlated forecast misses. For probability integral transform (PIT) evaluation, \citet{DieboldGuntherTay1998} show that marginal calibration can appear satisfactory while the joint dynamics are misspecified, a warning directly relevant to coupled systems where marginal supply and demand forecasts may look well-calibrated while the joint distribution is wrong.

For density forecasts, \citet{GneitingRaftery2007} establish the framework of proper scoring rules that reward sharpness subject to calibration. In multivariate settings, however, standard energy scores have limited sensitivity to misspecified dependence structures \citet{Pinson2013DiscriminationAO}. \citet{ScheuererHamill2015} propose variogram-based proper scoring rules that are explicitly sensitive to the dependence between forecast components. For coupled supply-demand forecasting, dependence-sensitive scores are essential because joint forecast errors propagate across sides.

Conformal prediction provides distribution-free coverage guarantees under exchangeability assumptions \citep{Romano2019,Barber2021}. For coupled systems, the challenge is that exchangeability is precisely what endogenous intervention violates. Developing conformal-style guarantees under the non-stationarity created by platform coupling is an open problem.

Recent work on tail calibration \citep{Allen2025} demonstrates that ``calibrated on average'' is insufficient for decision contexts dominated by tail risk. This is especially relevant when coupling amplifies extreme outcomes.

% ============================================================
\section{Search-to-booking distance metrics}
\label{app:distance}
% ============================================================

The search-to-booking gap (Section~\ref{sec:search_gap}) can be quantified by treating both the search lead-time distribution and the booking lead-time distribution as compositional objects on the simplex. Let $\mathbf{p}^{\text{search}}_t$ denote the search lead-time composition and $\mathbf{p}^{\text{book}}_t$ the booking lead-time composition in period $t$.

The Aitchison distance between the two compositions is
\[
d_A(\mathbf{p}^{\text{search}}_t, \mathbf{p}^{\text{book}}_t) = \| \text{clr}(\mathbf{p}^{\text{search}}_t) - \text{clr}(\mathbf{p}^{\text{book}}_t) \|_2,
\]
where $\text{clr}$ is the centered log-ratio transformation \citep{Aitchison1986}. The Aitchison distance is preferred for compositions because it is scale-invariant and subcompositionally coherent; note that the $\text{clr}$ transformation requires strictly positive components, so zero lead-time bins must be handled via small pseudocounts or multiplicative zero replacement before transformation.

Alternatively, the normalized $L_1$ distance used in \citet{KatzSavageColes2025} avoids the zero problem at the cost of losing the geometric properties of the Aitchison framework.

The distributional fitting framework of \citet{KatzNeedlemanMedina2026} provides the parametric machinery for modeling these compositions over time. The open problem is extending these tools to paired compositions and interpreting their divergence as a constraint indicator within a demand forecasting system.

Computing lead-time compositions requires internal search logs in which queries are tagged with intended stay dates (e.g., check-in/check-out fields or calendar interactions). Google Trends provides only a scalar index of search intensity over calendar time and cannot deliver lead-time distributions without strong additional assumptions; it is therefore suitable as a proxy for aggregate demand intent but not for the distributional gap measure.

% ============================================================
\bibliographystyle{plainnat}
\bibliography{references}

\end{document}